\documentclass[floats,twocolumn,amsfonts,amssymb,unsortedaddress,floatfix]{revtex4}
\input epsf
\usepackage{bm}
\usepackage{amsmath}
\usepackage{graphicx}
\usepackage{xcolor}
\usepackage{ulem}

\begin{document}

\def\Bbb{\mathbb}

\title[Trapped Colloids, Pinned Vortices and the Ice Rule]{Dumping Topological Charges on Neighbors: \\ Ice Manifolds for  Colloids and  Vortices}

\author{Cristiano Nisoli}

\address{Theoretical Division,  Los Alamos National Laboratory, Los Alamos, NM, 87545, USA}
\begin{abstract}

We investigate the recently  reported analogies between  pinned vortices in nano-structured superconductors or  colloids in optical traps, and  spin ice materials. The frustration of  colloids and vortices differs essentially from spin ice. However,  their effective energetics is made  identical by the contribution of an emergent field associated to the  topological charge, thus leading to a (quasi) ice manifold for lattices of even (odd) coordination. The equivalence extends to the local low-energy dynamics of the ice manifold, where the effect of geometric hard constraints can be subsumed into the spatial modulation of the emergent field, which mediates an entropic interaction between topological charges. There, as in spin ice materials, genuine ice manifolds enter a Coulomb phase, whereas quasi-ice manifolds  posses a well defined screening length, provided by a plasma of embedded topological charges. The equivalence between the two systems  breaks down in lattices of mixed coordination  because of  topological charge transfer between sub-latices. We discuss extensions to social and economical networks. 
\end{abstract}


\maketitle

\tableofcontents

\section{Introduction: (Artificial) Frustrated Materials} 

A  recent multidisciplinary effort in the creation  and study of artificial frustrated nano-materials~\cite{Brunner2002, Bohlein2012, Libal2006, Libal2009, Ray2013, Latimer2013, Trastoy2013freezing, Olson2012, Wang2006,  Nisoli2013colloquium, Ke2008,Li2010,  Ke2008a, Nisoli2010, Nisoli2007, Morgan2010, Morgan2013real, Nisoli2012, Ladak2010, Zhang2013,Porro2013, Zhang2011a, Qi2008,Kapaklis2012, Rougemaille2011, Branford2012, Farhan2013, Lammert2010, Morrison2013, Chern2013, Mellado2012, Han2008, Moller2006,Chern2013magnetic, Nascimento2012,Mol2009} has provided  mesoscale realizations accessible to direct visualization. These systems have led to the exploration of new of exotic states~\cite{Ladak2010, Branford2012, Zhang2013, Farhan2013,Gilbert2014}, including dynamics of magnetic charges and  monopoles~\cite{Castelnovo2008}.  While these results suggest possible applications  to superconductivity, sensing, and information storage, they more importantly  represent an attempt to design, rather than discover, desired emergent phenomena in the low-energy physics of interacting meso-structures, whose realization is now afforded by  recent advances in synthesis and integration.

Among these materials, the so called artificial spin ice~\cite{Wang2006, Nisoli2013colloquium} has now reached a certain maturity and has attracted physicists from a variety of backgrounds~\cite{Nisoli2013colloquium}. The system is composed of magnetically interacting nano structures that behave as single-domain macrospins. Their mutual geometric arrangement can be engineered and underlies potential emergence from their low-energy collective behavior. Thus, although devised initially~\cite{Wang2006} as a  mesoscale analogue of the  frustration of natural pyrochlore magnets~\cite{Ramirez1999}, artificial spin ice has then raised its own interesting issues. There is now growing awareness that this kind of program might lead to a bottom-up engineering of new magnets with desired emergent properties,  whose functional units could be delocalized monopolar degrees of freedom, rather than the localized dipolar magnetic domains of nature-given materials.

\begin{figure}[t!!!]
\begin{center}
\includegraphics[width=.9\columnwidth]{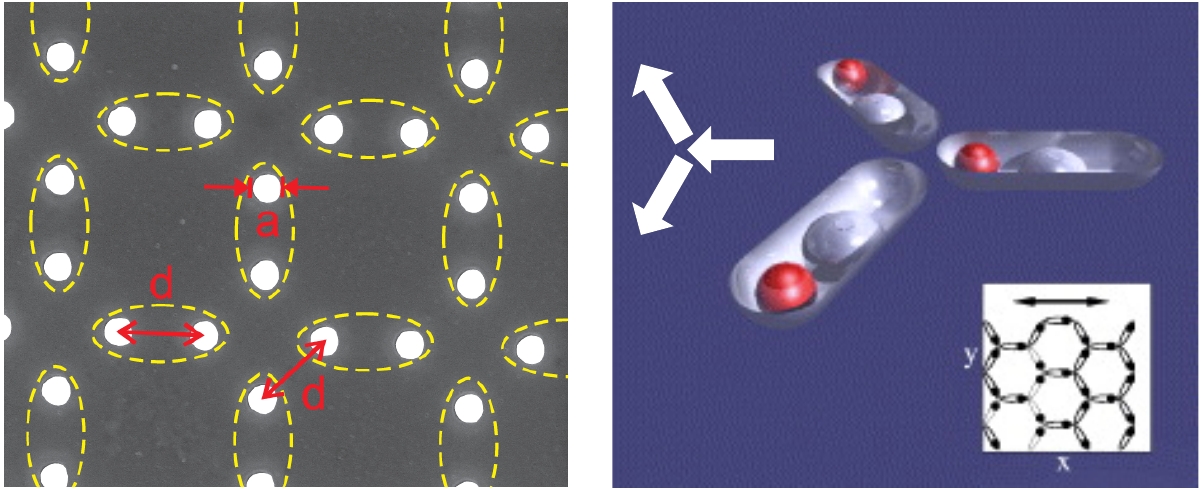}
\caption{Left: SEM image of the nano-patterned substrate for pinning of superconductive vortices in MoGe thin films, from~\cite{Latimer2013}. Right: Schematics of $z=3$, $d=2$ colloidal trap,  corresponding  hexagonal lattice, and spin equivalent, from~\cite{Olson2012}.}
\label{dis}
\end{center}
\end{figure}

However, proposals for realizations of frustrated analogues of spin-ice materials are not limited to magnetic systems. In  numerical works, Libal {\it et al.} proposed systems of colloids  held in place by elongated optical traps arranged along the sides of a two-dimensional lattice (typically square~\cite{Libal2006}, or hexagonal). Each trap has a double-well potential forcing the colloids to be  in the proximity of one of the two vertices connected by the trap (Fig.~\ref{dis}). When brownian dynamics is performed on these systems they obey the ice rule (for a square lattice) or quasi-ice rule (for an hexagonal lattice)  in the strong interaction regime. While the same ice rule is also observed in natural and artificial spin ice materials, its origin is different, as we shall see. 

The ice rule has also been proposed  by the same authors for quantum vortices ~\cite{Libal2009, Olson2012} in  properly nano-structured superconductors, and unsurprisingly so. Clearly the model is the same: whether colloids or vortices, we have in both cases certain objects subjected to hard constraints (a link with half occupation) and repulsive interactions, which are the strongest in the vertices of the lattice.  Some of these predictions were recently realized experimentally~\cite{Latimer2013,Trastoy2013freezing}. Naturally,
research  in pinning superconductive vortices to a substrate nano-patterned with holes  had been conducted since the late seventies~\cite{Fiory}, with the goal of  increasing the critical current. Latimer {\it {\it et al.}}, however, arranged the pinning with the intention of reproducing the frustration of an ice-like material in the allocation of vortices~\cite{Latimer2013}. They fabricated superconducting thin films of MoGe containing pairs of circular holes arranged as in Fig.~1:  a square lattice whose vertices can accommodate in principle four vortices each. Then, if the applied magnetic field is at half the matching value, only half of the holes are occupied. They found that at each vertex only two holes are occupied, and two are not, following the ice rule, as predicted numerically~\cite{Libal2009}.

These most interesting results need now to be placed on some more solid theoretical footing. First of all one immediately notices that  the models of Libal, Olson, Reichhardt and collaborators are over-constrained compared with the experimental realization of Latimer {\it et al.}, where there are no semi occupied link-shaped traps  and the vortices can simply pin to in the  ratio of vortices to hole dictated by the  field. But even more relevantly  it cannot be escaped that while the ice manifold appears both in colloids/vortex systems and in magnetic spin ice materials, its origin is rather different as different are the energetics involved. 

In this article we discuss  similarities and differences between these systems  and (artificial) spin ice. We investigate when, why, and how colloids/vortices can access an ice manifold, and when their ice-manifold disappears. We show how the collective behavior leads to  an effective energetics of the nodes that replicates the energetics of spin ice materials. 
This is obtained by including an emergent field conjugated to the  topological charge of the vertex. We then investigate  spatial modulations, and  we show that the equivalence extends to the low-energy physics above the ice manifold, where the emergent field mediates an entropic interaction between topological charges. Similarly to spin ice materials, the ice manifold (for lattices of even coordination) is in a Coulomb phase, whereas quasi-ice manifold (for lattices of odd coordination) possesses a finite screening length provided by its overall neutral plasma of embedded charges. Finally, because the similarity is consequence of topological charge conservation, we show that it breaks down in lattices of mixed coordination. There a net transfer of topological charge between differently coordinated nodes {\it must} occur---something inherently impossible in magnetic spin ice materials ~\cite{Morrison2013, Chern2013}. We also hint at extensions to networks and trees, on which we will report elsewhere. 

\section{Algebraic Constraint: Ice Manifolds} 

The ice rule is so named because in water ice each oxygen atom sits at the center of a proton sharing tetrahedron. Two protons are close and covalently bonded to the oxygen, whereas the two others are close to a neighbor. The resulting 2in/2out ice rule, proposed by Pauling to explain the experimentally reported residual entropy of water ice~\cite{Pauling1935} thus originates in the stoichiometry of water.   Famously, Pauling showed, the exponential freedom in choosing  configurations within the ice manifold allows  it to retain a finite density of residual entropy at very low temperatures. 
In spin ice materials (natural~\cite{Ramirez1999} or artificial~\cite{Nisoli2013colloquium}) protons are replaced by classical macrospins, and the ice rule   (2 spins pointing in, 2 spins pointing out for a $z=4$ coordination lattice~\cite{Ramirez1999, Wang2006, Nisoli2013colloquium}) or quasi-ice-rule  (1-in/2-out, and 2-in/1-out for $z=3$ lattices~\cite{Qi2008, Nisoli2010, Nisoli2013colloquium}) is dictated by minimization of the  frustrated  energies {\it of the vertices}.  

Thus, for (artificial) spin ice, it is the pairwise interaction between nearest neighboring spins converging in the same vertex that is frustrated. Indeed, if one describes these material in terms of a vertex-model~\cite{Lieb1967exact, Baxter1982}, as it is often done~\cite{Nisoli2007,Nisoli2010, Morrison2013, Morgan2013real}, then the ice-rule vertices have the lowest energy. Instead, the  frustration of colloidal systems presented by Libal {\it {\it et al.}} is of the emergent kind, a collective effect of the entire lattice.

In the following we will introduce a model to  describe  these systems most simply, analytically, and with some generality. In the experiment of Latimer {\it et al.}, where holes are nano patterned in a MoGe thin film, at half matching field, only half of the holes are pinning a vortices. This introduces an algebraic constraint on the total number of pinned vortices. With respect of this experimental reality, the systems typically studied by Libal, Olson, Reichhardt and collaborators however exhibit geometric hard constraints: each vertex of the lattice is connected by links, and each link contains only one colloid/vortex, which can occupy only the extremities. Obviously the geometric constraint implies the algebraic constraint, whereas the opposite is not true. In the experimental arrangement  e.g. it is not  impossible for two neighboring vertices to both harbor $z$ colloids~\footnote{Unfavorable energetics notwithstanding.} ($z$ being the coordination of the lattice), but the same is impossible in the numerical models considered by Libal and collaborators.

We will show  that the ice rule is consequence of the algebraic constraint, whereas later we will show that the hard constraint implied by links is responsible for emergent interactions between topological charges.

\subsection{Mean Field}

Consider  a lattice of vertices connected by links with coordination $z$. On each link sits a colloid (or a vortex) which can only occupy the extreme ends (Fig~\ref{dis}). This situation reproduces the system treated numerically in Ref~\cite{Libal2006, Libal2009, Olson2012}. The colloids (or vortices) considered in these works repel each other, via screened Coulomb (or modified Bessel) interaction.  We assume that these objects repel each other with an energy ${\cal E}$, and  we neglect interactions between colloids belonging to different nodes, thus treating the system as a vertex-model. (Numerical results have shown that the ice manifold emerges in regimes in which the screening length of colloids or the London penetration length is much smaller than the lattice constant~\cite{Libal2006, Libal2009}). Then, since each node can have $n=0, 1, \dots z$ close colloids, we can write
\begin{equation}
E_n= \frac{{\cal E}}{2} \,\, {n(n-1)},
\label{en}
\end{equation} 
for the energy of such configurations, each of multiplicity $m_n={z \choose n}$. The  reader might have already spotted an imperfection in our treatment. Eq. (\ref{en}) fails to describe  non-symmetric interactions in the vertex, which are typical for instance of the square geometry. We discuss this point later, and for the moment we ask the reader to bear with us.

The first thing to notice is that while our system is analogous to a spin ice material with spins directed along the links and pointing toward the colloid (Fig.~1), the energetics of (\ref{en})  
differ completely from the frustrated energetics of  spin ice. In fact, it is not frustrated at all: very simply, objects repel each other, and the lowest, and unfrustrated, vertex configuration corresponds to all the colloids (or all but one) being happily pushed away, and dumped on neighbors. However, such configuration of minimum energy   obviously cannot be achieved by all vertices (nor most vertices) of the lattice. Therefore, for colloids, not the pairwise interaction, but rather the allocation of all vertices  in the lowest  energy states is frustrated by the lattice. 

The frustration is thus of the emergent kind. In  spin ice, because of the protection afforded by time reversal symmetry,  this kind of emergent frustration in allocation of the vertex topology can only happen in dedicated and non-trivial geometries, such as those  introduced  by the author and collaborators~\cite{Morrison2013, Chern2013}, and recently realized experimentally~\cite{Gilbert2014}. Instead, in trivial geometries,   degeneracy in spin ice  follows from a pairwise frustrated energetics that is already degenerate {\it at the vertex level}: so much so that Pauling famous estimate of the residual entropy of ice~\cite{Pauling1935},  which applies just as well to the 2D ice~\cite{Baxter1982, Lieb1967exact} and hexagonal ice~\cite{Moller2009,Lammert2012}, was based solely on vertex degeneracy, and disregarded completely the mutual arrangement of vertices.  And  indeed in artificial spin ice the ice rule is accessed even by disjointed vertices and clusters, as shown experimentally~\cite{Li2010}, something that would  not happen for colloids on disjointed clusters of traps. 

In our current notation, a spin ice energetics  would imply that $n$ is degenerate with $z-n$,  which in magnetic materials follows from time reversal symmetry. Clearly this is not true for (\ref{en}). However the ice-rule obeying ground state ($n=z/2$ in all nodes if $z$ is even, or all nodes in $n=(z\pm1)/2$ if $z$ is odd) can  be easily inferred from the quadratic energetics of (\ref{en}). Yet, the similarity with spin ice  extends to non-zero temperatures: it is in fact possible to fold back this emergent frustration into a ice-like vertex energetics.

We will start with a mean field approximation that disregards any correlation between vertices and thus neglects the geometric constrains implicit in the analyses of Libal {\it et al.} and is more agreeable  to the Latimer experiment. However,  due to the disorder of the ground state, our initial conclusions  describe rather well the mean quantities in the geometric constrained systems.  

  We  introduce $\rho_n$, the probability of any node to  be in the  $n$-configuration.  The  ``free energy'' of an uncorrelated gas of nodes  is thus~\cite{Nisoli2010, Nisoli2007, Morgan2013real}
\begin{equation}
f=\sum_{n=0}^z\left( E_n \rho_n  +T \rho_n \ln \frac{\rho_n}{m_n}\right)-\kappa \left(\sum_{n=0}^z\rho_n-1\right),
\label{f}
\end{equation}
where  the entropy is chosen as the number of ways in which the distribution of vertex types can be realized, via any process that returns a consistent and unbiased statistical ensemble: the Lagrange multiplier $T$ can thus represent  an effective~\cite{Nisoli2007,Nisoli2010,Nisoli2012,Morgan2010, Morgan2013real} or real~\cite{Zhang2013, Porro2013} temperature; $k$ simply enforces normalization of the probability. 
 
Unlike in spin ice, to derive predictions directly from (\ref{f}) we need to enforce the algebraic constraint. Clearly, the inclusion of the  geometric hard constraints implied by Libal's systems  is a rather serious matter, with which we will deal later. However vertex-frustration follows in fact already from the simpler algebraic constraint present in Latimer's experiment: the  conservation of  colloids/vortices in the graph, which describes the  inability to collectively dump them on neighbors. To establish the similarity with spin ice materials, it is useful to introduce the topological charge associated to the configuration $n$. This is given by
\begin{equation}
q_n=2n-z,
\label{qn} 
\end{equation}
which is zero for the ice-rule ($n=z/2$): for spin ice $q_n$ is in fact the magnetic charge of the vertex. Then, the algebraic constraint dictates that any distribution $\rho_n$ must neutralize the average charge, or  ${\cal Q}=\sum^z_{n=0} q_n \rho_n$=0, which in turn  implies $\overline n=\sum^z_{n=0} n \rho_n=z/2$~\footnote{This  simply means that  the ice rule is  obeyed {\it in average}. It is not obvious that it should be obeyed at each vertex}. In fact we can request $ {\cal Q}+{\cal Q}^e=0$ if an excess charge ${\cal Q}^e$ is doped {\it extensively} into the system by adding  or subtracting colloids to the links, or by changing the magnetic field for vortices.~\footnote{We need  to open a brief parenthesis on the excess charge, which will be used in the following sections yet needs to be included in the formalism since the beginning (the reader can skip this paragraph at the first reading). Unlike in spin ice materials, where magnetic monopoles do not exist as elementary particles, here extra charge  can  be added or subtracted relatively easily.  In the case of hard constraints, if an extra colloid is added (or subtracted) to a link,  the link contains two (or zero) colloids rather than one, and is thus saturated (or empty). The situation does not appear as completely symmetric, since the two colloids in a link repel each other. If colloids are absolutely prevented to hop from link to link, the extra energy does not act on any of our degrees of freedom.  However, one can imagine that extra colloids can in fact hop---at least to a degree. When the extra energy from the inter-link interaction inside a saturated link is comparable to the energy barrier to escape the link-trap, a colloid might then hop out of the saturated link into a nearby link.  At the same time one can tune the traps such that a colloid in an half occupied link can hop into an empty link, while it cannot hop into another half occupied link because of the resulting repulsion from the  colloid already present. One sees how the system can be engineered to have either mobile, positive excess charges or mobile positive {\it and} negative excess charges, something inherently impossible in spin ice materials.  For pinned vortices in superconductors patterned with holes the situation is much easier: excess charge simply corresponds to tweaking the magnetic field around the half matching value. In particular in experimental realizations~\cite{Latimer2013, Trastoy2013freezing} there are no hard constraints and all the charge (not only excess charge) is always mobile. As we will see in following sections, extensive doping of charge can break the ice manifold; non-extensive doping simply  leads to local screening effects. In this section however we will consider ${\cal Q}^e=0$.}

To include the algebraic constraint, we  minimize
\begin{equation}
f_{\mathrm{tot}}=f+\phi\left(\sum_{n=0}^z q_n\rho_n+{\cal Q}^e\right)
\label{fT}
\end{equation}
where $\phi$ is at this point simply a Lagrange multiplier ensuring charge neutralization/conservation. For fixed
 $\phi$, minimization with respect to $\rho_n$ and $k$ returns the  usual Boltzmann distribution
\begin{equation}
\rho_n={z \choose n}\frac{\exp(- E^{\phi}_n/T)}{Z(T,\phi)},
\label{rho}
\end{equation}
 in the new, effective energies $E_n^{\phi}$ given by
\begin{equation}
E_n^{\phi}=E_n+q_n\phi.
\label{Enfi}
\end{equation}
Note that in (\ref{Enfi}) the value of the energy is offset  by what looks like an ``electrostatic'' contribution from   $\phi$, which  can then be considered  an emergent field  coupled to the topological  charge $q_n$. When dealing with the hard geometric constrain  and its effect on spatially modulated probabilities, we shall see that $\phi$ indeed deserves the title.

The partition function in (\ref{rho}) is given as usual by 
\begin{equation}
Z(T,\phi)=\sum^z_{n=0} {z \choose n} \exp(- E^{\phi}_n/T),
\label{Z}
\end{equation}
and  $\phi$ is found via   $\partial_{\phi}f_{\mathrm{tot}}=0$,  which can then be written as
\begin{equation}
{\cal Q}+{\cal Q}^e=-T\partial_{\phi}\ln Z(T,\phi)=0.
\label{Q}
\end{equation}

Solving (\ref{Q}) determines $\phi$ and therefore, through (\ref{Enfi}), also $\rho_n$ in (\ref{rho}), thus closing the problem. 
From (\ref{Q}), in general $\phi$  depends on $T$---which is indeed  always the case for graphs of mixed coordination, or when ${\cal Q}^e\ne0$ as we will see below.
However, for a lattice of single coordination and ${\cal Q}^e=0$ (no extensive doping), it is immediate to realize that the following temperature-independent choice 
\begin{equation}
\bar \phi=\frac{(1-z)}{4}{\cal E},
\label{fi}
\end{equation}
solves the problem of ensuring charge neutralization at an temperature. Indeed, from (\ref{fi}) and (\ref{Enfi}) we find  
\begin{equation}
 E_n^{\bar\phi}= \frac{E_n+E_{z-n}}{2}=\frac{{\cal E}}{8}\left[q_n^2+z(z-2)\right].
\label{Enfibar}
\end{equation}
The last equality in (\ref{Enfibar})  establishes  an ice-like energetics in the absolute value of the topological charges,  therefore ensuing ${\cal Q}=0$ at any temperature. Thus the collective effect of the lattice is subsumed into a vertex description, courtesy of an emergent field that offsets the purely interactive energy by taking into account the constraint  of charge  conservation.

It is now useful to  relabel nodes in terms of  charge $q$ from (\ref{qn})  rather than colloids $n$. From (\ref{Enfibar}) and (\ref{rho}) we have $\rho_q=\rho_{-q}$.
\begin{figure}[t!!!]
\begin{center}
\includegraphics[width=.9\columnwidth]{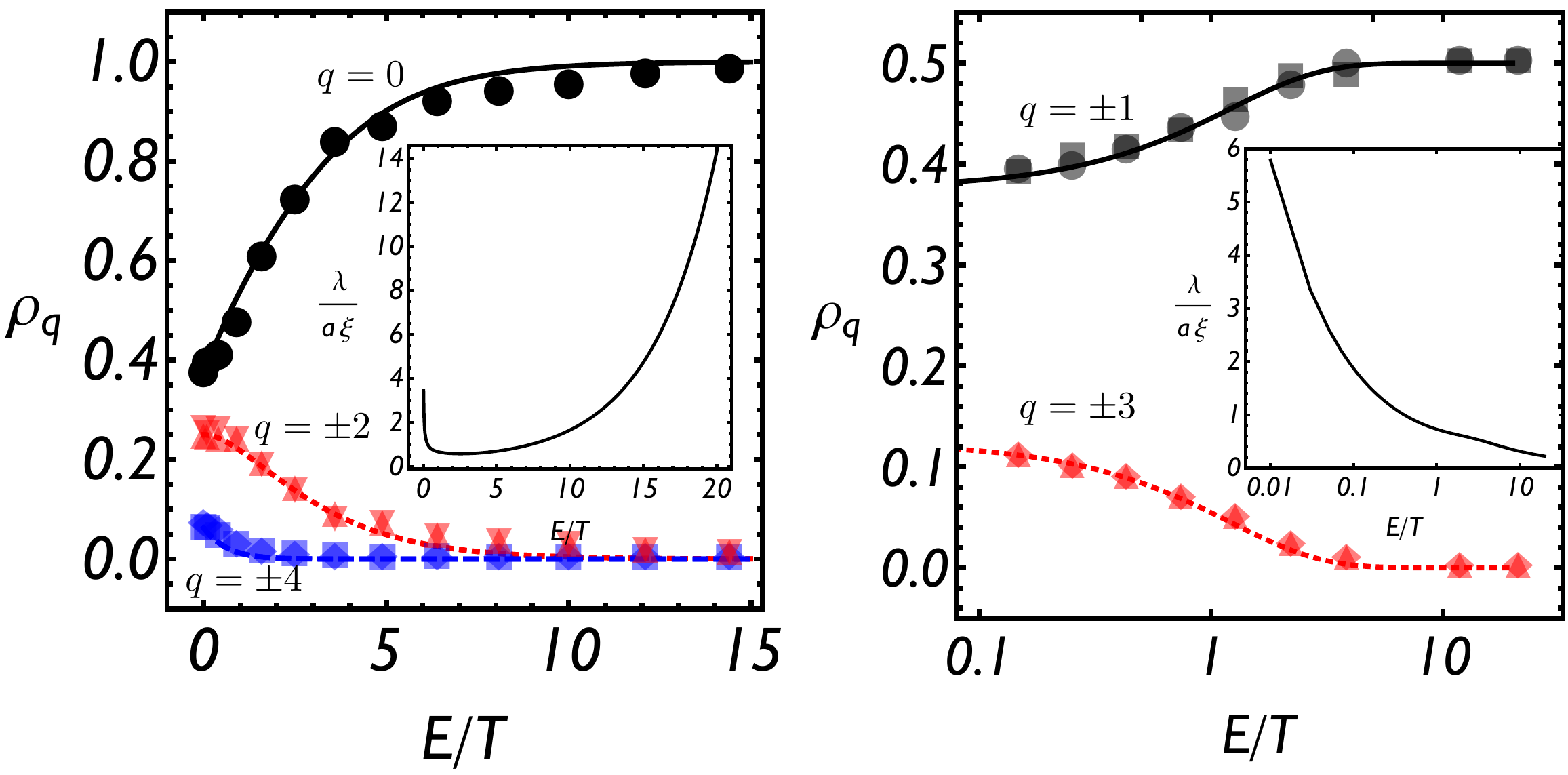}
\caption{Left:  For $z=4$, relative vertex frequencies $\rho_q$ as a function of $E/T$ as in (\ref{rho}) for $q=0$ (black, solid), $q=\pm2$ (red, dotted), $q=\pm4$ (blue, dashed), are plotted against numerical data for the square colloidal lattice obtained by Libal {\it et al.}~\cite{Libal2006}, for vertex populations corresponding to $n=2$, $q=0$ ($\bullet$), $n=1$, $q=-2$ ($\blacktriangle$), $n=3$, $q=2$ ($\blacktriangledown$), $n=0$, $q=-4$ ($\blacklozenge$), $n=4$, $q=0$ ($\blacksquare$). Right: for $z=3$ relative vertex frequencies $\rho_q$ as a function of $E/T$ as in (\ref{rho}) for  $q=\pm1$ (black, solid), $q=\pm3$ (red, dotted) are plotted agains numerical data for the hexagonal superconductive vortex system of Libal {\it et al.}~\cite{Libal2009}, for vertex populations corresponding to $n=1$ ($\bullet$), $n=2$($\blacksquare$), $n=3$ ($\blacktriangle$), $n=0$ ($\blacklozenge$). Insets: the screening length as a function of $E/T$ showing the exponential divergence (left)  corresponding to the ice-manifold.}
\label{data}
\end{center}
\end{figure}
%
From this energetics, at low temperature the system crosses over to a manifold containing only the vertices of lowest charges. 
For  even  $z$, for which vertices of $q=0$ are allowed when $n=z/2$, the system
enters a  low temperature, genuine ice manifold (Fig.~\ref{data}). For odd $z$ the lowest effective vertex-energy pertains to charges $q=\pm1$, corresponding to $n=(z\pm1)/2$. Then at low temperature the system enters a  {\it quasi}-ice manifold: as the odd coordination prevents charge cancellation at the vertex level, the quasi-ice manifold is a plasma of embedded  charges $q=\pm 1$,  represented  in equal proportion~\cite{Libal2009, Olson2012}. 


\subsection{Fits of Numerical Results}

Although our results were obtained simply by imposing the algebraic constrain of topological charge cancellation, interestingly they can be used to fit the strongly constrained model of Libal {\it et al.}  The authors of these works considered two cases, the hexagonal and square lattice.

Applications involving 2-D hexagonal lattices~\cite{Olson2012} ($z=3$, $d=2$) fall into this framework. In Fig.~\ref{data}, right panel, we plot data obtained from  simulated brownian dynamics of a superconductive vortex system (Fig. 4.b in Ref. \cite{Libal2009})  vs. the predictions of (\ref{rho}). The data pertains to different lattice constant measured in multiples of the London penetration depth $\lambda$. As the effective ${\cal E}/T$ for the repulsive interaction of vortices scales with the the modified Bessel function $K_1(a/\lambda)$, where $a$ is the lattice constant, we plot the frequencies obtained numerically by Libal {\it {\it et al.}} {\it vs.} $\alpha K_1(a/\lambda)$, where $\alpha$ is a fitting parameter. We found very good agreements with our predictions for $\alpha=950$. Black squares (circles) represent frequencies of vertices with charge $q=+1$, ($q=-1$) or $n=2$ ($n=1$). Red triangles (diamonds) represent frequencies of vertices with charge $q=+3$, ($q=-3$) or $n=3$ ($n=0$). Black (red) lines are predictions for charge $q=\pm1$ ($q=\pm3$). 

The case of a square lattice is trickier and forces us to finally discuss the limitations of (\ref{en}). It is well known both in colloidal systems~\cite{Libal2006}, vortices in supercunductors~\cite{Latimer2013,Trastoy2013freezing} and artificial spin ice~\cite{Morgan2010, Zhang2013} that  the degeneracy within  the ice manifold is lifted  in 2D square lattice realizations ($z=4,~d=2$), because of a difference in distance and therefore interaction strength between colloids or spins impinging in the same vertex. This leads to low temperature ordering within the ice manifold. This ordering  can be included  in our approach,  by properly tinkering (via symmetry breaking within the ice rule topologies) with  the energetics in (\ref{en}). However, considering the scope of our current analysis, to proceed here along that route would not be particularly illuminating. Indeed the interesting physics of this ordering pertains to the formation of domain walls in the ordered phases---which can have two different  ``orientations''~\cite{Libal2006, Morgan2010}---and  the out of equilibrium issues related to their motion, which has been used to address more general and quite interesting issues of grain boundaries dynamics. However, these predictions on local behavior would escape our mean field approach anyway. Moreover our scope is to analyze similarities between ice systems, and therefore we concentrate  on models that can access a degenerate and thus genuine ice  manifolds rather than ordered phases.  Relevantly, there are many ways in which the degeneracy of the ice manifold can be maintained even in a square lattice, for instance the proposal of M\"oller~\cite{Moller2006} as well as others, still unpublished, yet discussed in the artificial spin ice community. 

Even disregarding the possible symmetry breaking within the ice-manifold, our treatment still accounts for the average frequency of ice-rule vertices.  Fig.~\ref{data}, left panel, demonstrate this  by plotting data points for Brownian dynamics of colloids (Fig 2.a in Ref.~\cite{Libal2006}), grouped by topological charge, {\it vs.}  ${\cal E}/T$. Black circles denote  ice-rule vertices, of charge $q=0$ and thus $n=2$. Red triangles pointing down (up) represent frequencies of vertices with charge $q=+2$, ($q=-2$) or $n=3$ ($n=1$). Blue squares (diamonds) represent frequencies of vertices with charge $q=+4$, ($q=-4$) or $n=4$ ($n=0$). Solid lines are predictions obtained from (\ref{rho}) for  $q=0$ (black), $q=\pm2$ (red) $q=\pm4$ (blue). 

We have shown that although the frustration in trapped colloids is of the emergent kind and thus differs essentially from that of spin ice materials, their behavior can be seen as controlled, at least in a mean field treatment, by a similar, effective energetics, which adds to the actual energetic  the contribution of an emergent field $\phi$ associated to the topological charge.  This (quasi) ice rule is consequence of the algebraic constraint of charge cancellation and we will show in Section 4. that it is lifted by charge transfers. In the next section we will show that in the case of hard geometric constrained provided by links, the analogy can be pushed to the low energy dynamics of the ice manifold.

\section{Geometric Constraint: Emergent Interactions} 

Lest the reader grows rightfully annoyed at the author for calling emergent field what is simply an uniform Lagrange multiplier, we show how $\phi$ can provide information on the low-energy dynamics and local perturbations,  as a field should. 

\subsection{Spatial Modulation}

For a spatially varying probabilities, the intuitive generalization of (\ref{rho})  should allow for a non uniform field $\phi$ in (\ref{Enfi}), and thus non uniform probabilities $\rho_n(x)$, in the form
\begin{equation}
\rho_n(x)={z \choose n}\frac{\exp\{- [E^{\bar \phi}_n+q_n\phi(x)]/T\}}{Z(T,\phi)}.
\label{rhox}
\end{equation}
Then a field equation is needed for $\phi(x)$. Let us see how that can come  across more formally.

If $\rho_n(x)$ is the probability of a node  $x$ to be in configuration $n$, then the  free energy $f$ in (\ref{f}) generalizes to the functional
 \begin{equation}
 {\cal F}[\rho]=\sum_x f(\rho(x))+\Delta {\cal F}[q],
 \label{F}
 \end{equation}
that adds to the uncorrelated local free energy (\ref{f})  the non-local term $\Delta{\cal F}[q]$, which accounts for the effect of the underlying spin structure, including  charge conservation and correlation effects. 
The  reader will have noted that we have already introduced a low temperature approximation in (\ref{F}), since $\Delta{\cal F}$  depends on $\rho_n(x)$ through  the density of charge $q(x)=\sum_nq_n\rho_n(x)$.  
To fathom the form of $\Delta {\cal F}[q]$ we introduce $\phi$ as the conjugate field
\begin{equation}
\phi(x)=\frac{\delta \Delta {\cal F}}{\delta q(x)}
\label{jn}
\end{equation}
 which leads to  the  Legendre transform
\begin{equation}
{\cal L}[\phi]=\left(\Delta{\cal F}-q \cdot \phi \right)_{q=q[\phi]}
\label{L}
\end{equation}
[where $q \cdot \phi= \sum_{x} q (x)\phi(x)$], which implies
\begin{equation}
q(x)= -\frac{\delta{\cal L}}{\delta \phi(x)}.
\label{qfi}
\end{equation}
and thus finally
\begin{equation}
{\cal F}[\rho,\phi]=\sum_x f(\rho(x))+q \cdot \phi+{\cal L}[\phi].
\label{F2}
\end{equation}
The local part of the functional in (\ref{F2}) (first two terms on the right) looks now more like the uniform average free energy in (\ref{fT}) and the non-local functional ${\cal L}[\phi]$ (third term) pertains to the emergent  field, which mediates an entropic interaction. 

We can now construct ${\cal L}$ by perturbing over our previous uncorrelated treatment. We saw in the previous section that the uniform formalism based solely on the algebraic constraint could predict well the numerical results for the average vertex frequencies. Therefore $\rho_n$ in ($\ref{rho}$), the probability of {\it any} vertex to have $n$ closed colloids,  should be given by  $\rho_n=N_v^{-1}\sum_x \rho_n(x)$ ($N_v$ is the number of nodes). 

It follows that our functionals, restricted to uniform fields, should reduce to the previous forms of (\ref{fT}). 
From (\ref{jn}),  when $\rho_n(x)$ are uniform, so is $\phi(x)$. Then, in order to recover (\ref{fT}) from (\ref{F2}), 
${\cal L}$,  {\it restricted to  uniform fields}, must be
\begin{equation}
{\cal L}[\phi]=\sum_x{q^e(x)} \phi =N_v {\cal Q}^e \phi
\label{Lunif}
\end{equation}
[$q^e(x)$ is the excess charge in the node $x$ 
and ${\cal Q}^e=N_v^{-1}\sum_x q^e(x)$ is the average excess charge per node]. 

 \begin{figure}[t!!!]
\begin{center}
\includegraphics[width=.85\columnwidth]{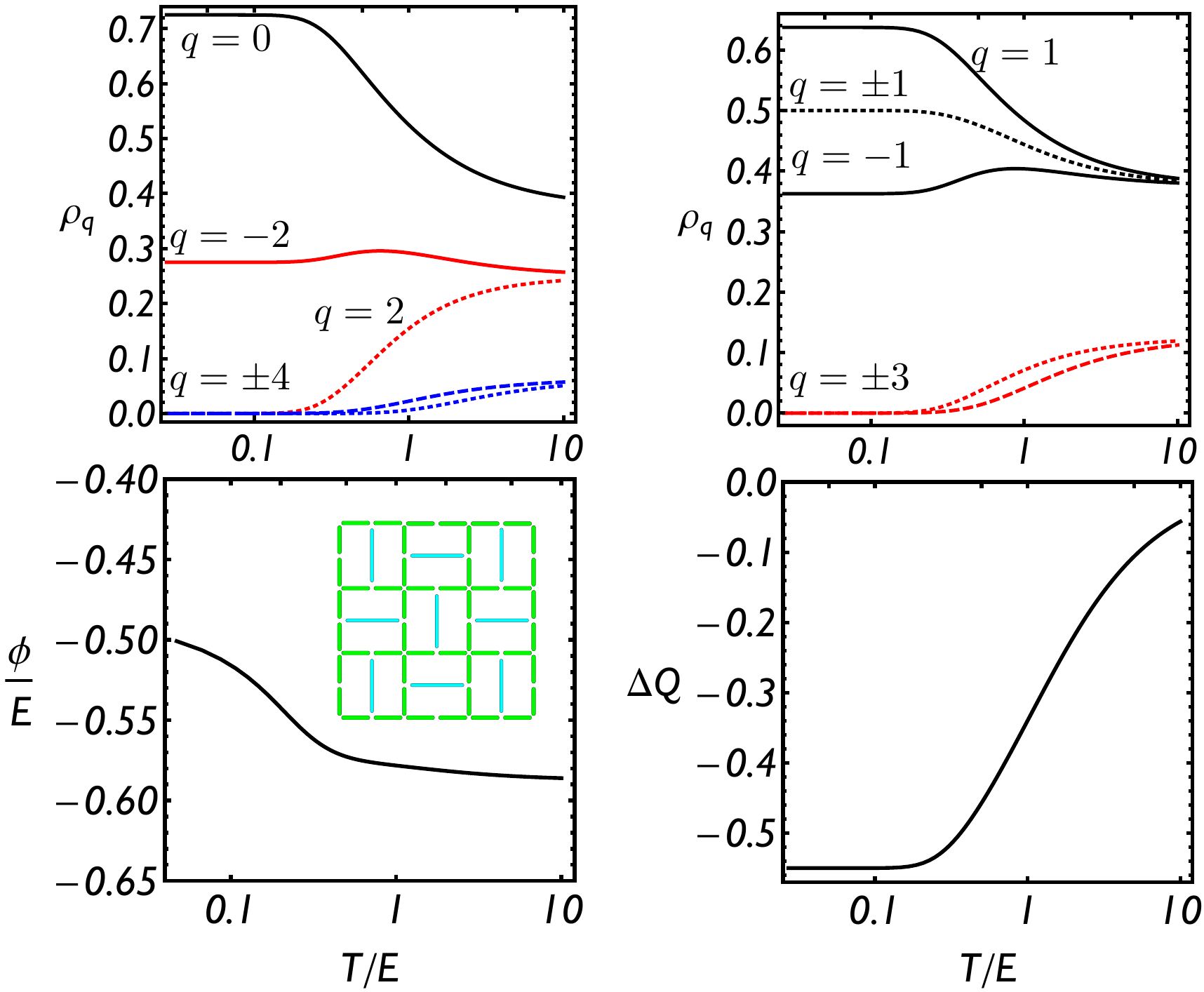}
\caption{For a lattice of mixed coordination there {\it must} be a net transfer of topological charge between vertices of different coordination (bottom right). For $z=3,4$ (inset: the shakti lattice of~\cite{Morrison2013, Chern2013}) $\phi$ at low $T$ approaches its value for $z=3$, corresponding to $E(1-z)/4=-E/2$ (bottom left): thus the sub-vertices of $z=3$  obey the quasi-ice rule (top right, the $q=\pm1$ line, black dotted, is split), but with an imbalance of positive $q=1$ charge, to screen the excess charge coming from the $z=4$ vertices (top left).  }
\label{Mixed}
\end{center}
\end{figure}

Perturbing over the uniform, average manifold we expand in the derivatives of $\phi$.~\footnote{
This perturbative approach to weak correlation, viable on a disordered manifold, would be impossible in an ordered phase such as the one obtained by lifting the degeneracy within the ice manifold for the square colloidal case.} We assume that  the lattice is regular and allows coarse graining of $x$ into a continuum variable, and thus $\sum_x\to a^{-d}\int_{L^d} d^dx$, where $a^d=L^d/N_v$ is the volume of the unit cell.  At second order the only admissible form is
\begin{equation}
{\cal L}[\phi]=\int_{L^d} \left[q^e \phi -\frac{1}{2}\epsilon \, \partial_i\phi \partial^i \phi\right]\frac{d^dx}{a^d}.
\label{L2}
\end{equation}
Indeed, to  reduce to (\ref{Lunif}) for uniform fields, (\ref{L2}) must contain only terms at second order in the derivatives of $\phi$, thus excluding terms such as $\phi^2$ or $ \partial^i \phi \partial _i q^e$. Here $\epsilon$ is  the generalized permittivity of the emergent field $\phi$ (in general one has  $\epsilon_{ij}$, a suitable  tensor). 

\subsection{Entropic Debye Screening}

The solution can now be obtained by optimizing the functional (\ref{F2}) in the fields $\rho_n(x)$ and $\phi(x)$.
In taking the functional derivative with respect to $\phi$  we cannot dump its derivatives at the boundaries since we already know that $\phi$ is not zero at infinity: indeed its spatial average must be $\bar \phi$. It is convenient to replace $\phi(x)\to \bar \phi+\phi(x)$, with $\phi \xrightarrow{x\to \infty}0$, and  minimize in both. Minimization in $\phi(x)$ then returns
\begin{equation}
-\Delta\phi=(q+ q^e)/\epsilon,
\label{coulomb}
\end{equation}
which  from (\ref{coulomb}), (\ref{L2}), and (\ref{F2}), also gives
\begin{equation}
{\cal F}=\int_{L^d} \left[f+ \epsilon \frac{1}{2}\partial_i \phi\partial^i \phi\right]\frac{d^dx}{a^d}
\end{equation}
and thus  $\epsilon>0$.  

Then optimization  of (\ref{F2}) with respect of $\rho(x)$  returns the Boltzmann law  (\ref{rhox}), whereas optimization with respect of $\bar \phi$ leads  again to the charge constraint ${\cal Q} +{\cal Q}^e=0$  for the spatially modulated $\rho_n(x)$  of (\ref{rhox}).
We have obtained  a Debye-H\"ukel model for an electrolyte solution (not uncommon in spin ice materials~\cite{Castelnovo2011})where  charges are topological while the interaction $\phi$  is emergent from the   underlying  network of links: the excess charges are screened by the  charges of the manyfold.

Consider ${\cal Q}^e=0$ but $q^e(x)\ne0$\footnote{Note that ${\cal Q}^e=0$  does neither imply absence of excess  charge $q^e(x)$, nor that the net excess charge $Q^e\int_{L^d} q^e(x)d^d x/a^d$ is zero, or even finite, but only that it is sub-extensive. Then  the {average excess charge per node} ${\cal Q}^e=Q^e/N_v$ is zero.}. Expanding  (\ref{rhox}) in $\phi(x)$ around $\bar \phi={\cal E}(1-z)/4$, one finds $\rho_n(x)=\rho_n+\eta_n(x)$, with  $\rho_n$  given by (\ref{rho}) and $\eta_n$ given by
\begin{equation}
\eta_n(x) =- \rho_n  q_n  \phi(x)/T
\label{xibar}
\end{equation}
which correctly implies $\sum_n \eta_n=0$. Then from (\ref{xibar}) and (\ref{coulomb}) and the hypothesis that the excess charge is sub-extensive (${\cal Q}^e=0$ ), we have $\rho_n=N_v^{-1}\int_{L^d}\rho_n(x)d^dx/a^d$: $\rho_n$ is the probability of {\it any} vertex to be in topology $n$, as expected. 

Since $\sum_n\rho_n q_n=0$, we have $q(x)=\sum_n\eta_n(x) q_n$ and from (\ref{xibar})
\begin{equation}
 q(x)=-\overline{{\cal Q}^2} \phi(x)/T
\label{q}
\end{equation}
where $\overline{{\cal Q}^2}=\sum_n \rho_n q_n^2$ is the average  charge fluctuation of the  manifold. Finally from (\ref{coulomb}) and (\ref{q}) $\phi$ satisfies 
\begin{equation}
(\lambda^{-2}-\Delta)\phi=q^e/\epsilon,
\label{phi}
\end{equation}
a screened Poisson equation whose screening length 
\begin{equation}
\lambda=\sqrt{{\epsilon \, T}/{\overline{{\cal Q}^2} }},
\label{lambda}
\end{equation}
 precisely corresponds to the  Debye formula where the embedded screening charge comes from the charge fluctuation within the manifold.

  Since  a genuine ice manifold  (even $z$) lacks  charge fluctuation, from (\ref{lambda}) we have $\lambda^{-1}=0$ and infinite screening length.  A non-extensive excess charge does not disturb a  genuine ice manifold: indeed from  (\ref{q}) since $\overline{ {\cal Q}^2}=0$ we have   $q(x)= 0$ and therefore all the vertices stay in an ice configuration.  This can also seen in (\ref{xibar}) which gives, in all cases,  $\eta_n (x)=0$  in the ice manifold, and therefore $\rho_n(x)=\rho_n$: the ice manifold remains uniform in presence of excess charge, which must therefore being pushed at the boundaries. Indeed $\lambda$ approaches infinity exponentially fast in ${\cal E}/T\to \infty$, as depicted in Fig~\ref{data}, because $\overline{ {\cal Q}^2} \propto \exp(-{\cal E}/2T)$. 
Therefore as the system crosses over into an ice manifold it  approaches an entropic solenoidal (or Coulomb) phase  for $\phi$ in (\ref{phi}). Then standard potential theory in any dimension implies that a mobile excess charge is expelled at the boundaries as the system enters the ice rule. At small or non-zero temperature mobile excess  charges will have a density $\lambda^{-d}$ which goes to zero at $T$ approaches zero.  

If the charge is not mobile equations (\ref{xibar}) still predict $\eta_n(x)=0$. Consider an ice manifold. Now add an extra colloid to saturate a link, causing a charge +2 in a vertex. The ice manifold is restored by flipping ``spins'' to move the charge to the infinitely distant boundaries, thus creating a "Dirac string" from the bulk to the boundaries. Then $q(x)$ remains zero in the bulk, as expected. However if ${\cal Q^e}\ne0$ and the excess charge is extensive, this cannot happen (see next section). 

In absence of excess charge, excitations at $T>0$ interact via (\ref{coulomb}), as the magnetic monopoles~\cite{Castelnovo2008} of spin ice, yet with a difference: because of the short range energetics hereby assumed~\cite{Libal2006}, the interaction between monopoles  is here entirely of the emergent kind and dimensionality dependent: for $d=3$ it is a Coulomb potential and opposite charges are separable, while for $d=2$ they are logarithmically confined.

Conversely in a quasi-ice manifold (odd $z$) there are always embedded $\pm1$ charges and thus $\overline {{\cal Q}^2}=1$. This  plasma of embedded charges always provide for entropic  screening of   excess charges, something seen numerically in colloidal systems~\cite{Libal2009}, but also in artificial spin ice of odd coordination~\cite{Chern2013, Chern2013magnetic}. As temperature lowers this screening becomes tighter and can form bound states, or polarons~\cite{Chern2013magnetic}, which, when  spaced at a distance much larger than $\lambda$, simply fluctuate thermally. 
 
As temperature increases, one expects the screening to become less tight. Indeed, consider now $q^e=0$. When $T/{\cal E}\to \infty$  all links flip independently and vertices are allocated by multiplicity.  From (\ref{rhox}) this implies $\phi/T \xrightarrow{T/{\cal E}\to \infty} 0$, which from (\ref{coulomb}) entails $\epsilon T \xrightarrow{T/{\cal E}\to \infty} \infty$.  Since $\epsilon$ is inversely proportional to an energy,  dimensional considerations fixes it at $\epsilon=\xi^2 a^2 {\cal E}^{-1}$ where $\xi$ is a number, and is limited in ${\cal E}/T$. Then (\ref{lambda})  implies, correctly, $\lambda^2/a^2\propto T/E$ (since $\overline{{\cal Q}^2}\le \overline{{\cal Q}^2}|_{T=\infty}=z$). 

In conclusion, while a uniform emergent field enforces the algebraic constraint and lead to the ice rule, a modulated emergent field mediates the effect of the geometric constraint via a local entropic interaction between charged excitations and excess charges, explains the attraction of opposite topological charges seen in numerical simulations~\cite{Libal2009}, and extends the equivalence with magnetic spin ice materials  to the low-energy dynamics above the ice manyfold. 

\section{Charge Transfer: Breaking the Ice}

Where does the similarities between repulsive colloids/vortices and spin ice break down? We have seen that the ice rule follows from the algebraic constraint of charge cancellation. On a lattice of single coordination this constrained can be violated by extensively doping excess charge. However in a lattice of mixed coordination, even without excess charge, a charge transfer between sub lattices of different coordination can lift the ice manifolds. 

\subsection{Extensive Doping}

As explained above, unlike in spin ice materials, excess charge (colloids) can here be added or subtracted easily to the systems under considerations.
If the doping is non-extensive (${\cal Q}^e=0$)   effects are only local, leaving the previous picture unchanged. Conversely, adding extensively a (possibly  negative) average number $n^e$ of colloids induces an average topological charge per unit vertex ${\cal Q}^e=-2n^e$. This charge breaks the ice---but not the quasi ice---manifold.  

 If $n^e\ll1$  
(only a small fractions of all the vertices are defected) 
we can apply  the previous approach with ${\cal Q}^e\ne0$ in (\ref{Q}). Now $\phi$ can depend on $T$. However, from  (\ref{rho}) and (\ref{Enfi}),   $\rho_n$ are determined by the expansion 
\begin{equation}
\phi(T)=\phi_0+\alpha T +O(T^2),
\label{fiexp}
\end{equation}
of $\phi$  at the first ordering $T$: 
 $\phi_0$ determines the energetics in (\ref{Enfi}), and can be chosen to make either one  or two states  degenerate; $\alpha$ simply adds a factor  to the multiplicities , $m_n \to m_n\exp( -q_n\alpha)$, and thus controls the ratio of degenerate states at low $T$.

In lattices of odd $z$, which at low temperature enter a quasi-ice phase of embedded charges $q=\pm1$ in equal proportion,  the excess charge can be neutralized within the  manifold. Then $\phi_0=-(z-1){\cal E}/4$ as in (\ref{fi}) and $\alpha$  fixes the multiplicities of the $q=\pm1$ charges, such that 
\begin{equation}
\rho_{q=1} \xrightarrow{T\to 0}(1-{\cal Q}^e)/2, ~~~~~~\rho_{q=-1} \xrightarrow{T\to 0}(1+{\cal Q}^e)/2.
\label{qeodd}
\end{equation}
We see therefore that the presence of embedded charges in the quasi-ice manifold allows for neutralizing of a small density of external charge without leaving the manifold but simply by changing the relative proportions of the charges. 
 To make sense of (\ref{qeodd}): consider a quasi-ice manifold, for instance $z=3$. Each vertex is in the $n=2$, $q=+1$ or $n=1$, $q=-1$ configuration. 
 If we add a  charge ${\cal Q}^e$,  half of the excess colloids (or $|{\cal Q}^e|/4$ colloids) goes on  $n=1$, $q=-1$ vertices turning them in $n=2$, $q=+1$.  Half  fall on $q=+1$, $n=2$ vertices creating $n=3$ excitations over the manifold. These will dump the excess charge via a different, unsaturated link, on a neighboring  vertex in the $n=1$, $q=-1$ topology (or else non-neighboring,  though a cascade effect), thus eliminating the excitation and  regaining  the $n=2$ status.  
 
The situation is rather different for a genuine ice manifold, which contains no charge of its own: then extensive excitations   
are needed to absorb the excess charge. Assume ${\cal Q}^e$ positive. We can always chose $\phi_0$ so that the $q=-2$ charges (corresponding to $n=z/2-1$) are degenerate with the $q=0$ charges (corresponding to $n=z/2$) in the energetics of (\ref{Enfi}). Then $\alpha$ in (\ref{fiexp}) gauges the relative admixture of the two giving
\begin{equation}
\rho_{q=0}\xrightarrow{T\to 0}1-{\cal Q}^e/2, ~~~~\rho_{q=-2}\xrightarrow{T\to 0} {\cal Q}^e/2
\label{qeeven}
\end{equation}
or the screening of the excess charge by means of a suitable density of $q=-2$ excitations. Dropping colloids (or subtracting them) on a ice-rule manifold in an extensive, diluted way  creates  excitations in the very vertex in which the colloid falls, excitations that  cannot be suppressed by  flipping unsaturated links (as explained in the previous section). 

This difference in behavior between odd and even coordination number is relevant to lattices of mixed coordination, where a transfer of net topological charge eliminates the ice manifold.

\subsection{Mixed Coordination} 

Consider lattices of mixed coordination number, an intriguing scenario that opens a window on more complex geometries~\cite{Morrison2013, Chern2013, Chern2013magnetic} and  in general on dynamics in complex networks, which we will develop elsewhere. Then the free energy 
 is the sum of terms given by (\ref{f}), each corresponding to sub-lattices of different coordination, and weighted by the relative abundance of vertices of that coordination. However, the emergent field must be the same for all sub-lattices. Indeed it is the total charge, not the sub-lattice charges, that must be conserved. 
 
Consider for simplicity the treatment of section two,  which neglects the geometric hard constraints. One finds that in each sub-latice  the probability of a node of coordination $z$ to be in topology $n$ follows a Boltzmann distribution as in (\ref{rho}), however the field $\phi$ must be the same for all $z$.   This implies that there {\it must be}  charge transfer  between sub-lattices of different coordination,  since (\ref{fi}) cannot be satisfied for all $z$ simultaneously by the same field. Therefore   at most  one sub-lattice can reach the ice (or quasi-ice) manifold at low $T$, whereas the others  are no longer equivalent to a spin ice system. 
This situation is distinctively different from the case of artificial spin ice of mixed coordination, which instead  always enters an ice manifold~\cite{Morrison2013, Chern2013}: there, charge conservation is implied by an  energetics that is genuinely  degenerate in the sign of the charge.

For definiteness, consider the case of  mixed coordination  4 and 3, in Fig.~\ref{Mixed}.  Our previous discussion on doping  and (\ref{fiexp}) can be employed. If we choose $\phi_0/E =-1/2$ in (\ref{fiexp}), then the $z=3$ sub-lattice  enters the quasi-ice manifold at low $T$. Yet in the $z=4$ sector we have from (\ref{Enfi}) that $q=-2$ and $q=0$ become degenerate and of lowest effective energy. This means that the $z=4$ vertices  dump positive topological charge on the  $z=3$ ones by  ``exciting'' negative charges ($q=-2$). 

The reason is straightforward. In truth, the $q=-2$ configurations are  excitations only in the {\it effective} energetics (\ref{Enfi}) for a lattice of single coordination number. For mixed coordination they are are in fact a way to lower the {\it real} energy in (\ref{en}), by pushing away colloids whenever geometry permits. Then the $z=3$ sub-lattice  can screen the excess charge without abandoning the quasi-ice manifold, in a way reminiscent of what happens to the shakti artificial spin ice~\cite{Morrison2013, Chern2013, Gilbert2014}  and pentagonal~\cite{Chern2013magnetic} artificial spin ice {\it above} their ice manifold. 
  
\subsection{Extensions to Economics}

This framework, suggested by physics, invites intriguing extensions to sociological and economical problems. Frustration being, after all, a typical aspect of our lives, one see how the  treatment above can be generalized naturally to  sociological networks of actors sharing burdens, undesirable tasks or costs, and how it can be employed to predict  optimal distributions of connectivity in a social setting. If ${\cal E}$ is negative, however, the model describes instead a network of shared benefits, or desirable wealth. In both cases net wealth (or cost) only amounts to half the available slots---which we can call opportunities in the case of wealth (${\cal E}<0$), or availabilities in the case of burdens (${\cal E}>0$).  Then the treatment of section 2 describes a completely fluid situation, in which costs or benefits can go freely from a node to another. 

We will show in future work that this lack of hard geometric constraints leads trivially to  extreme polarization of wealth:  actors with more opportunities have all their opportunities satisfied whereas authors below a certain opportunity threshold simply get nothing. The case of shared costs is intrinsically more fair: above a certain availability threshold everybody performs the same number of tasks,  regardless of their availabilities. Below that threshold, performance equals availability. In both cases the ice rule is broken. 

We clearly see how this facile picture is then severely modified by the introduction of hard constraints. These correspond to joining in a pairwise way the opportunities/availabilities of different actors through links of shared ``colloids'', thus obtaining a network generalization of the lattices described above. For instance, by coordinating among themselves actors with large opportunity, and thus forcing them to share wealth, one breaks down the wealth disparity. Conversely we can cement disparity by assuring that highly coordinated actors have enough links with lower coordinated ones, allowing the former to draw wealth from the latter. 

To maintain the physical focus on the current work, we will  offer a  treatment of these sociologically relevant models in a future report. There we will also show ways to account for the hard constraint in a mean field fashion and we will test them on exact solutions for peculiar graphs generalizations of Bethe lattice.

\section{Conclusion} We have studied the equivalence between systems of trapped colloids or pinned vortices,  and  spin ice. We have seen that although their energetics differs essentially from that of the spin ice systems, a constraint on topological charge cancellation introduces an ice-like effective energetics. The energetics is altered by a uniform emergent scalar field associated with the topological charge. Because of that,  lattices of even coordination number can access an ice manifold. Lattices of odd coordination  access a quasi-ice manifold. In the case  of strong geometrical constraints, the analogy extends to the low energy dynamics. Then a spatially modulated emergent field translates the geometric constraint into an entropic interaction. In genuine ice manifold such interaction is solenoidal, leading to a coulomb phase. In lattices of odd coordination, which access a quasi-ice-rule characterized by embedded charges, the charge fluctuations of the manifold can screen the excess charge, as also seen in artificial spin ice, in which polarons can form. Finally the equivalence breaks down in lattices of mixed coordination, whose behavior is essentially different from mixed coordination spin ices. In the case of colloids charge transfer must occur between vertices of different coordination, thus lifting the ice rule. 

\section{Acknowledgments} We thank P. E. Lammert for  critical reading of the manuscript, C. \& C. Reichhardt and A. Libal for providing  numerical data. Figures are reprinted with permission from the authors. This work was carried out under the auspices of the National Nuclear Security Administration of the U.S. Department of Energy at Los Alamos National Laboratory under Contract No.~DEAC52-06NA25396. 

\newpage

\bibliography{library.bib}{}

\end{document}